# Gold nanoparticles generated in ethosome bilayers, as revealed by cryo-electron-tomography

Patricia de la Presa,[*,†] Tatiana Rueda,[†] María del Puerto Morales,[‡] F. Javier Chichón,[§] Rocío Arranz,[§] José María Valpuesta,[§] and Antonio Hernando[†]

*Instituto de Magnetismo Aplicado, UCM-ADIF-CSIC, P.O. Box 155, Las Rozas, Madrid, 28230 Spain, Instituto de Ciencia de Materiales de Madrid, CSIC Cantoblanco, 28049 Madrid, Spain, and Centro Nacional de Biotecnología, CSIC Cantoblanco, 28049 Madrid, Spain*

E-mail: pdelapresa@adif.es,Tel:+34913007173,Fax:+34913007176

---

[†]Instituto de Magnetismo Aplicado
[‡]Instituto de Ciencia de Materiales de Madrid
[§]Centro Nacional de Biotecnología

**Abstract**

Gold nanoparticles have been synthesized inside ethosomes, vesicles composed of phospholipid, ethanol and water, which could be very efficient not only in delivery probes to the skin but also as diagnostic and therapeutic multimodal agents. High efficiency encapsulation of gold nanoparticles is achieved by a simple strategy: the nanoparticles synthesis occurs simultaneously with the ethosomes formation, in the absence of any undesirable reducing agents. A three-dimensional reconstruction of a gold-embedded ethosome generated by cryoelectron tomography reveals that the gold particle is localized inside the lipid bilayer, leaving the ethosome surface and core free for further functionalization. The resulting gold nanoparticles are homogeneous in size and shape and, depending on synthesis temperature, the size ranges from 10 to 20 nm, as revealed by TEM. The ethosome-nanoparticles hybrids size has been investigated by means of dynamic light scattering and has been found to vary with temperature and gold salt concentration from 700 to 400 nm. Gold nanoparticles encapsulated ethosomes offer a versatile platform for the enhancement of pharmacological efficacy in transdermal and dermal delivery systems.

# Introduction

Liposomes are good candidates for controlled reactions, and they have been used as microreactors to synthesize metal oxides,[1] or metallic nanoparticles (NPs).[2,3] The simplicity of the liposome synthesis method has an important advantage over other encapsulation methods such as polyelectrolyte capsules synthesized by layer by layer adsorption.[4] Furthermore, liposomes encapsulating some of these NPs offer great opportunities in terms of nanoscale delivery system as therapeutic or diagnostic multimodal agents.[5–7] Particularly, liposomes containing gold NPs can be used as probes to follow biological processes through microscopic techniques such as liposome internalization and transport inside cells or blood vessels or even to increase the UV light-induced release of drugs.[8]

As therapeutic agents, light absorbing NPs -as gold- are able to kill cells upon exposure to short laser pulse.[9] The localized surface plasmon resonance of gold NPs has a strong optical absorption in the UV-visible spectra, which enables the development of diverse biomedical applications.[10] For diagnostic, the resonance frequency can be tuned to the near-infrared, allowing the NPs to act as molecular contrast agents in a spectral region where tissues are transparent. The localized heating due to resonant absorption, enables also thermal ablation therapies and drug delivery mechanisms. These properties together with their general lack of toxicity, suggest that noble metal NPs encapsulated liposome are highly promising nanomaterials for novel biological applications.[11]

The synthesis of gold NPs in uni- or multilamellar vesicles have been already developed by other authors.[12–15] Chow et al.[14] reported on the synthesis, in the presence of a reductant, of liposomes with a significant amount of gold nanoparticles, but rather inhomogeneous in shape and size. Meldrum et al.[13] showed that the reduction of Au(III) can take place in the absence of an external agent, only under the presence of the lipid molecules egg yolk phosphatidylcholine (PC). However, the amount of Au NPs per vesicle was very low, as shown by transmission electron microscopy (TEM) images. Regev et al.[15] reported the spontaneous, in-situ synthesis of gold NPs within onion-type multilamellar vesicles without any additional chemical reductant, since the vesicle component, the monoolein, can act as such. Although isolated gold NPs are obtained by this method, they are polydisperse both in size and shape, ranging from 8 to 105 nm, and forming flat epitaxed triangular crystals with truncated corners or sphere-like or even elongated particles.

In this paper, gold NPs have been generated for the first time in the lipid bilayers of ethosomes,[16] vesicles composed of phospholipid, ethanol and water. These capsules are soft, malleable vesicles which are able to generate biocompatible NPs and to encapsulate and deliver through the skin highly lipophylic molecules as well as cationic or hydrophilic drugs.[17–19] Ethosome encapsulated gold NPs offer then a versatile platform for the improvement of pharmacological efficiency.

Experimental conditions are tailored to synthesize uniform gold NPs within the lipid bilayers of unilamellar ethosomes with hydrodynamic size as small as 400 nm. Gold nanoparticles in terms of size, degree of aggregations and aggregate shape have been investigated by classical TEM micrograph, whereas the colloidal suspension of ethosomes-nanoparticles hybrid has been analyzed by dynamic light scattering and negative staining, which allowed as a first approach to locate gold nanoparticles in ethosomes. However, this last technique has limitations due to the impossibility of localizing three-dimensional objects in the two-dimensional projection of a micrograph. To solve the problem, we have used cryo-electron microscopy that allows the direct visualization of the vesicles in a fully-hydrated environment, maintaining the ethosome shape and structure unaltered, and allowing the use of tomographic, three-dimensional reconstruction techniques. As far as we know, this is the first time that this powerful technique has been used for the characterization of ethosomes-nanoparticles hybrids.

# Experimental section

## Chemicals

Hydrogen Tetrachloroaurate (III) trihydrate, 99.99%, $HAuCl_4.3HO_2$ and Lecithin (Phosphatidylcholine), Refined 99% were purchased from Alfa Aesar. Ethanol Absolute 99.5% ($CH_3CH_2OH$) and Trichloromethane (chloroform) 99% ($CHCl_3$) were purchased from Panreac. All reagents were used without further purification. Water was purified with a Pobel Water Still 3-4 l/h water purifier.

## Synthesis

The strategy of encapsulating gold in ethosomes consists basically on mixing Phosphatidylcholine (PC), ethanol and gold salt $AuClH_4$ under mild heating conditions. Different parameters are involved in the synthesis of the NPs, such as temperature, PC and gold salt concentration, water and ethanol ratio, which determine the nanoparticle size and the morphology of ethosome-nanoparticles hybrids. Therefore, different samples have been synthesized in order to understand the influence of each parameter on the formation of gold nanoparticles and on the ethosome encapsulation. First, a blank of ethosome was synthesized as described by Touitou et al.[16] (Figure 1(A)). Ethosome encapsulated gold NPs were prepared as follow: A solution containing 2-6% PC, 30-45% ethanol and 2-20 mM was mixed and slowly heated at a temperature range from 33 to 68 °C. The stirring was maintained during 30 min to homogenize the sample. During this time, no change in the color was observed. After that, deionized water to 100% w/w was added, and the color slowly begun to change from pale yellow to purple, characteristic of small gold NPs formation. After 30 min the solution was completely purple and homogeneous, the intense reddish-purple color indicative of the generation of isolated small gold particles (Figure 1(B)). A similar procedure was followed to prepare a mixture of PC, $HAuCl_4$ and water (Figure 1(C)). Finally, ethanol was replaced by chloroform (Figure 1(D)) producing gold NPs in micelles.

## Characterization

### Dynamic Light Scattering

We performed dynamic light scattering (DLS) measurements in a ZETASIZER NANO-ZS device (Malvern Instruments Ltd, UK) to determine the hydrodynamic size of the ethosomes-nanoparticles hybrids in a colloidal suspension. The samples have been previous diluted in the appropriate medium with 30 to 45% ethanolic solution in order to avoid multiple diffusion effects that reduce the hydrodynamic radius and increase the ratio signal-noise.

**Particle size characterization**

Particle size was determined from TEM micrographs in a 200 kV JEOL-2000 FXII microscope. For the observation of the sample in the microscope, a drop of the suspension was placed onto a copper grid covered by a carbon film and was allowed to evaporate. The mean particle size, $d$, and the logarithmic standard deviation, $\sigma$, were obtained from digitized TEM images by counting more than 100 particles.

**Negative stain electron microscopy**

Samples were negatively stained with a solution of 2%(w/v) uranyl acetate and visualized using a JEOL 1200 EXII electron microscope operated at 100 kV and at a nominal magnification of 15000.

**Cryo-electron-tomography**

Samples were applied to holey-carbon grids (Quantifoil® R 3.5/1) and plunged into liquid ethane in a Leica EM-CPC vitrification device. The grids were used for tomographic acquisition under low dose conditions on a FEI TecnaiG2 electron cryomicroscope operating at 200 kV. The tilt-series were acquired with the FEI Xplore3D© software at a nominal defocus of -5 to -8 $\mu$m using a Saxton scheme (from -65° to -65°, normally a set of 57 to 67 images). Images were recorded with a 794 Gatan SlowScan 1k x1k CCD camera at a nominal magnification of 14500 with a pixel size of 1.23 nm/pixel.

The tilt-series were processed in a Linux Fujitsu-Siemens CeliusV810 bi-opteron with 16 Gb of RAM. The 16 bit MRC image formatted stacks were pre-processed and aligned using the IMOD[20] software package using the gold nanoparticles as fiducials.[21] The aligned tilt-series were low-pass filtered to 20 Å and reconstructed using 8 iterations of the iterative TAPIR (Tomographic Alternating Projection Iterative Reconstruction) algorithm implemented in the IVE/Priism software package. TAPIR uses an iterative real-space method similar to SIRT (Simultaneous Iterative Reconstruction Technique) to compute the tomogram. In each iteration, the tomogram is refined by back-projection of the residual difference between the experimental projections in the tilt series and the calculated projections from the tomogram, while applying nonlinear spatial constraints. The spatial constraints, which include positivity (i.e., negative density values are set to zero) and the boundedness (i.e., the reconstructed density is confined to the specimen thickness), provide better robustness against the artifacts due to the missing wedge.[22]

For each reconstructed volume, a denoising procedure to discern and interpret the structural features within the volumetric representation was applied. To this end, Gaussian filtering[23] (standard deviation set to 0.5) and anisotropic nonlinear diffusion (AND) were combined using TOMOAND software.[24,25] Denoised volumes were segmented using AMIRA® (TGS Europe, Merignac, France) with semi-automatic masking generation tools. For the visualization of the three-dimensional volumes both AMIRA® and Chimera 24 softwares was used.[26]

# Results and discussion

Figure 2 shows a typical view of the formed gold NPs, which are rather homogeneous in shape and size, the latter depending on the synthesis temperature whereas gold salt concentration affects the NPs cluster size.

Since different parameters are involved in the synthesis of the NPs, the influence of each one by varying one per turn is investigated. First, keeping the PC (4%), $HAuCl_4$ (10 mM), ethanol (40 %, w/w) and water ratio constant, the synthesization temperature is varied in order to observe the influence on the size of NPs and on the agglomerate of ethosome encapsulated gold NPs.

It is observed that the size of the NPs decreases as the temperature increases (see Figure 3). Because the NPs are not totally spherical, the maximum Feret's diameter is used to compute the size, i.e., the maximum perpendicular distance between parallel lines which are tangent to the perimeter at opposite sides. Fitting a lognormal curve to the size distribution results in an average nanoparticle diameter of 20 nm and $\sigma = 0.10$ (n = 122 particles) when the synthesis temperature is 33 °C. As the temperature increases to 68 °C, the mean diameter changes to 8 nm with $\sigma = 0.25$ (n = 139 particles) (see Figure 4). The decrease in the NPs sizes is due to the kinetic of the reaction:

as the temperature increases the gold atoms nucleate faster thus reducing the size of the NPs.

It is observed that the NP size is not affected by PC or gold salt concentration. However, the shape and aggregation degree of the gold NPs are significantly affected mainly by the amount of $HAuCl_4$. Figure 5 shows two kinds of clusters for gold concentrations 20 and 2 mM. For large gold concentrations, the NPs tend to form multipod clusters, whereas at low concentration the NPs are well dispersed and isolated. Thus, as the salt concentration increases, the NPs tend to form bunches inside the ethosome, whereas for smaller $HAuCl_4$ concentration, there are only one or two NPs per ethosome, which make them being separated by hundred of nanometers. It must be highlighted the large amount of gold NPs which can be loaded in the ethosomes.

Significant differences are observed for the sample synthesized with chloroform instead ethanol. Figure 6(A) and (B) show the TEM images of gold NPs synthesized in the ethosomes and in the chloroform, respectively, for large gold concentration (20 mM $HAuCl_4$), 4% PC, 35% ethanol-(chloroform) and synthesis temperature 43 °C. As can be seen in Figure 6(A), the gold NPs synthesized in the ethosomes tend to form bunches separated by several hundred of nanometers. On the other hand, when ethanol is replaced by chloroform, the gold NPs are well dispersed.

An interesting point is to determine the mechanism involved in the formation of the particles. The first question is whether an ethosome can act as microreactor, similarly to liposome microreactor of the metal oxides,[1] or metallic NPs[3] synthesis. To elucidate this question, the synthesis was carried out using chloroform, which is less polar than ethanol. In this case, micelles were formed with gold NPs in the apolar phase (outside the micelles) as revealed by TEM of negatively stained images of the specimen (see Figure 7). This result is in good agreement with the TEM image of gold NPs in chloroform (Figure 6(B)) which shows well dispersed gold NPs. The negatively stained TEM images confirm also that gold NPs are hydrophobic since they are located outside the polar phase in the hidrophobic oil phase. In a similar procedure, mixing $HAuCl_4$ (10 mM), PC (4%) and water (96 %) at 40 °C resulted in the formation of gold particles, but no ethosomes were generated. Therefore, the microreaction process in the ethosomes is discarded.

The second question is the reduction process involved, whether the PC acts as reducer or cata-lyst. The formation of gold clusters by chemical reduction of HAu$^{III}$Cl$_4$ has been well described by Gachard et al.[27] To produce Au(0) species from Au(III) the reduction Au(III) ions to Au(II) is required. The gold reduction may occur either by UV-irradiation or by chemical reduction. To evaluate the effect of light on gold reduction the synthesis was performed in the dark, and gold nanoparticles were generated, which ruled out light as a reducing agent. Recently, Regev et al.[15] have proposed that the lipid monoolein, the onion supplier of OH groups, acts as a reductant for the gold salt: R−CH−(OH)−CH$_2$(OH) being oxidized to R−CH(OH)−CO$_2$H when Au(III) ions are reduced to Au(0). On the other hand, it is known that ethanol as well as water can act as very weak reducer agents. To test this, 5 mM HAuCl$_4$ was mildly stirred in ethanol at 35 °C for a couple of days, but no changes in the color were observed. However, as the temperature was increased to 68 °C the color changed immediately from yellow pale to pink, indicating that the gold salt was completely reduced and the atoms nucleated forming small gold NPs. Therefore, considering that the synthesis was carried out at temperature lower than 45 °C, the ethanol can also be discarded as reductant at the working temperature. In a similar experiment, 5 mM HAuCl$_4$ was mildly stirred in water at 35 °C for one day. Unlike what happens with ethanol, smooth changes in color were observed after one day, indicating that water is a weak reductant at the working temperature.

It can be concluded that water is the reducer agent and that it is necessary for the formation of gold NPs, while external reducing agents or ethosome formation are not required, the PC acting as catalyst of the reduction. It is well known that gold has an enormous reducing facility. The methods can be very simple, such as the reduction of gold acetate by water and oleylamine (which also acts as a surfactant),[28] or sophisticated, like the Brusts method.[29] Depending on the reduction method, the gold NPs can have different sizes, shapes, can be widely functionalized, can have different optical properties and, as it has been recently shown, can even have magnetic properties.[30] In the present case, reduction of gold ion Au(III) to Au(0) -driven by PC in water- occurs in a milder manner and, consequently, the NPs are homogeneous in size and shape.

The analysis in a DLS experiment allows the determination of the hydrodynamic radius, i.e., the diameter of the ethosomes-nanoparticles hybrids in a colloidal suspension. Fixing PC, HAuCl$_4$ and ethanol to 4%, 10 mM and 40% respectively, the hydrodynamic size of the encapsulated NPs ethosomes decreases from 700 to 400 nm as the temperature increases from 30 to 35 °C (see Figure 8), then remaining constant for higher temperatures. The gold salt concentration also affects the hydrodynamic size in a similar way. Figure 8 shows that Dh decreases from 800 to 400 nm as the $HAuCl_4$ concentration increases from 2 to 5 mM (for 4% PC, ethanol 40% w/w and T = 43 °C).

To determine the location of gold particles in the ethosome, TEM was used. As a first approach, a negative staining traditional technique revealed the direct association of the gold particles to the ethosomes. Figure 9(A) and (B) are the corresponding TEM images of two different samples synthesized with 4% PC, at T = 35 °C and (A) 5 mM and (B) 1 mM of $HAuCl_4$. The agglomeration of the NPs is also similar to that observed by the TEM images of gold NPs (Figure 5): as the gold concentration increases, the gold NPs tend to form agglomerates (see Figure 9(A)), whereas for low gold concentrations they are isolated and separated by hundred of nanometers (see Figure 9(B)). However, in both cases the NPs are clearly associated to the ethosomal vesicles. Comparing the synthesis performed with ethanol (ethosomes, Figure 9) with that with chloroform (micelles, Figure 7), it is possible to observe significant differences: in Figure 7 all Au NPs are in the oil phase of the micelles, whereas in Figure 9 the NPs are inside the two-dimensional projection of the ethosomes. Although no gold NPs are observed outside the ethosomal vesicles, this method cannot unambiguously detect whether the particles are inside or outside the ethosomes, due to the indetermination produced by the fact that an electron micrograph is a two-dimensional projection of a three-dimensional specimen.

To solve this problem, cryo-electron-tomography[31] was used to locate the forming gold particles in 3D. This technique has definitive advantages over classical TEM since it allows the direct visualization of the vesicles in a fully-hydrated environment, which maintains the shape and structure of the ethosomes, and therefore allows the use of tomographic reconstruction techniques. Figure 10 shows a cryo-TEM image of the NPs encapsulated ethosomes, the arrows pointing to the gold NPs. The statistical analysis of the ethosome size from cryo-TEM images gives an average diameter of 140 nm with $\sigma = 0.25$ (n = 142 particles), which is considerably smaller than the values obtained by DLS. This is an artifact due to the experimental conditions. In cryo-TEM, only the thinner areas with small ethosomes (50-200 nm) properly embedded in the ice were choosen for the measurements. The larger Au NPs encapsulated ethosomes (diameter longer than 300 nm) form darker areas in which the ethosome scattering is not possible to visualize. From analysis of the cryo-TEM images, 46% of the ethosomes are unilamelar, 38% are bilamelar and an 17% contains more than two layers. Bilamelar ethosomes look actually like ethosomes piled up, without tightly apposed membranes, typical for the "onion" aspect of the multilamelar ethosomes. The bilamelar and the unilamelar ehosomes make up for 83% of the total number.

Very low concentration of gold was used (1 mM) to obtain one NP per ethosome in order to minimize the scattering artifacts produced by the gold within the tomographic reconstruction (Figure 10 and Movie1 in Supplementary Information (SI)). This artifact is not present in the projection images (Figure 11(A)) and appears in the tomographic volume as unresolved cone-shape areas perpendicular to the tilt axis (Figure 11(B); see also Figure1 and Movie2 in SI).

The reconstructed tomogram of a gold-containing ethosome (Figure 11(C) and Movie2 in SI) reveals that the gold particle is inserted into the ethosome bilayer. However, the area surrounding the gold particle cannot be reconstructed due to high scattering power of the metal cluster (Figure 11(B)). This problem leaves open two possibilities for the location of the gold particle within the ethosome membrane, either inserted into the bilayer but protruding from it and being exposed to the polar environment (Figure 11(D)) or embedded into the lipid bilayer and interacting with the internal sides of the two monolayers (Figure 11(E)). The high hydrophobicity of the gold particles and the flexibility of the lipid bilayers clearly points to the second possibility as the most likely one.

The results suggests a high encapsulation efficiency of the gold NPs in the ethosomes and a large stability of the NPs-ethosomes hybrids, specially for low gold concentrations. The formation of gold NPs is observable by eye inspection: the change in the color from yellow pale to deep reddish is indicative of the formation of small gold NPs. The amount of water and PC are enough to reduce all the gold salt presents in the dispersion. As shown by the TEM negative staining and cryo-TEM images, all the gold NPs are associated to the lipid bilayers of the ethosomes; furthermore, no precipitation of materials is observed up to a given time. Therefore, we assume that the gold encapsulation efficiency could be high, although further measurements must be carried out in order to quantify the encapsulation efficiency.

When the gold salt concentration is larger than 10 mM, precipitation of gold NPs is observed after one month. By shaking the dispersion it is possible to homogenize it. However, the NPs precipitate again after one hour. This is indicative that hydrophobic gold NPs have precipitated outside the lipid bilayers. For lower gold concentration, the stability of the dispersion increases: for concentration as low as 2 mM, it has be found more than one year stability.

Potential applications of the Au NPs encapsulated ethosomes are the localized heat therapies, for example hyperthermia. When colloidal gold NPs are administered to the blood circulation, the most important problem is their short circulation time in the bloodstream. The ethosomal system contains soft phospholipid vesicles in the presence of high concentrations of ethanol, which are able to overcome the natural dermal barrier, the stratum corneum. The encapsulation of the gold NPs in the lipid bilayer of ethosomes will allow the introduction and localization of gold NPs deep in the skin, and moreover, the NPs will remain protected by the ethosomes. As mentioned in the introduction, Au NPs can be heated by absorption of light, whereby the absorbed light energy is converted into thermal energy, heating locally the zone of interest. The Au NPs encapsulated ethosomes are then a promising material for therapies which requires local heat, as for example hyperthermia for skin cancer.

## Conclusion

The self-assembly of PC into organized structures results in enclosed molecular templates for the encapsulations of gold NPs. In the conventional methods employed for particle synthesis in liposomes, the internal concentration is not controlled, and separation steps are necessary to discard extravesicular NPs. Significant improvements of the method described here for gold-encapsulated ethosomes show that i) the NPs are homogeneous in size and shape, and that the size can be tuned by the temperature synthesis, but it is always lower than 25 nm, ii) the ethosomes are mostly unilamellar and the NPs lay inside the lipid bilayer, leaving the surface and the internal volume free for further functionalizations, and iii) the ethosomes are able to encapsulate a large amount of gold NPs; however, the lower the gold content is, the stabler the NPs encapsulated ethosomes are.

## Acknowledgement

Financial support from the Spanish MCyT under Project No. MAT2002-04246-c05-05 and CM under Project No. S-0505/MAT/0194 are acknowledged. PP acknowledges support from the Spanish Ministry of Education and Science through the Ramon y Cajal program. The work by JMV was supported by grants BFU2007-62382/BMC from the Spanish Ministry of Education and S-0505/MAT/0283 from the Madrid Regional Government. FJC, RA and JMV acknowledge to Jose López Carrascosa for his support thoughout the complection of this work.

## Supporting Information Available

Supporting Information Available: Cryo-TEM images and reconstructed volume of gold NPs encapsulated ethosomes. This material is available free of charge via the Internet at http://pubs.acs.org.

# Figure Legends

**Figure 1:** Sample colors for different syntheses: A) the ethosomes blank, B) the ethosome encapsulated Au NPs, C) a mixture of PC, HAuCl$_4$ and water, and D) gold NPs in micelles of PC, chloroform and water.

**Figure 2:** TEM image of gold NPs with a mean size of 20 nm. The scale bar represents 20 nm.

**Figure 3:** TEM images of gold NPs synthetized at 33° (left) and 68° (right). The scale bar represents 50 nm.

**Figure 4:** Size distribution of the NPs with the lognormal fitting curve. Left: distribution for synthesis temperature 68 °C. Right: synthesis temperature 33 °C.

**Figure 5:** TEM images of gold NPs for (A) large (20 mM) and (B) low (2 mM) $HAuCl_4$ concentrations (PC 4%, ethanol 40%). The scale bar represents 100 nm.

**Figure 6:** TEM image of gold NPs for 20 nm $HAuCl_4$ and 35% (A) ethanol. The scale bar represents 500 nm. (B) Chloroform. The scale bar represents 100 nm.

**Figure 7:** Negative stain micrograph of gold NPs in the micelles synthesized in PC, water and chloroform. The gold NPs are in the oil phase outside the micelles, confirming the hydrophobic nature of gold NPs surfaces. The scale bar represents 500 nm.

**Figure 8:** Hydrodynamic size for ethosomes-nanoparticles hybrids as function of temperature and gold concentration.

**Figure 9:** TEM negative stain micrographs of the ethosomes-nanoparticles hybrids for 4% PC, 30% w/w ethanol and (A) 5 mM and (B) 1 mM of $HAuCl_4$. The arrows point the Au NPs. Gold NPs are not observed outside the ethosomes. The scale bar represents 500 nm.

**Figure 10:** Cryo-electron image of the ethosome-nanoparticles hybrids. The arrows point the Au NPs. The asterisks mark crystals of water or ethane with a less electrondense aspect. The scale bar represents 500 nm

**Figure 11:** Cryo-electron-tomography and proposed models of the ethosome-nanoparticles hybrids. (A) 0° tilt image of a ethosome containing a gold particle. The scale bar represents 50 nm (B) Virtual slice from the reconstructed denoised volume of the same ethosome represented in (A). (C) 3D volumetric representation of the tomogram. A mesh recreates membrane in the unresolved area around the gold particle. (D) Possible model of the hybrid with the particle exposed to the internal and external ethosome environment. (E) Possible sandwich model of the hybrid with the particle protected by the hydrophobic sides of lipid bilayer.

# Legends for supporting material

**SIFig1**: Cryo-electron tomography from encapsulated gold NPs. (A) Projection image (0 ° tilt) from an ethosome with associated gold NPs (upper panel). The YZ (medium panel) and the XZ (lower panel) tomographic planes show the association of the gold NPs to the ethosome membrane. In the right panels, the ethosome membrane was manually segmented in red. (B) Projection image (0 ° tilt) of a small ethosome with an encapsulated gold-nanoparticle (upper panel). The XY (medium panel) and the YZ (lower panel) tomographic planes show the association of the gold nanoparticle to the ethosome membrane. The scale bars represent 50 nm.

**Movie1**: Volumetric representation of a reconstructed volume of nanogold-particles encapsulated in ethosomes. The density of the ethosome membrane is represented in blue whereas the gold density is in yellow. The gold NPs are directly associated to the ethosome membrane.

**Movie2**: The tomographic planes of the data move along the Z axis showing the ethosome structure and the encapsulated gold NP. The segmented membranes appear colored in blue as the rendered volume comes into view during the movement of the tomographic planes along the Z axis. A mesh represents the unresolved area near the gold particle. Finally the surface is clipped by a XY showing the position of the gold NP in the line described by the mesh as shown in Figure 11(C).

Figure_1

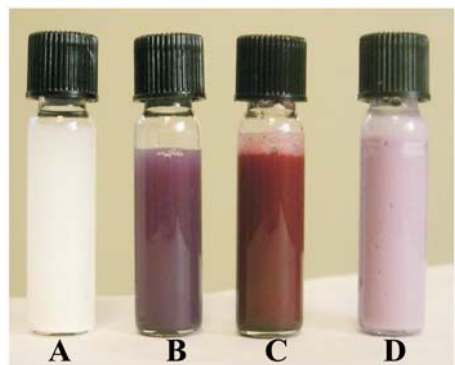

Figure_2

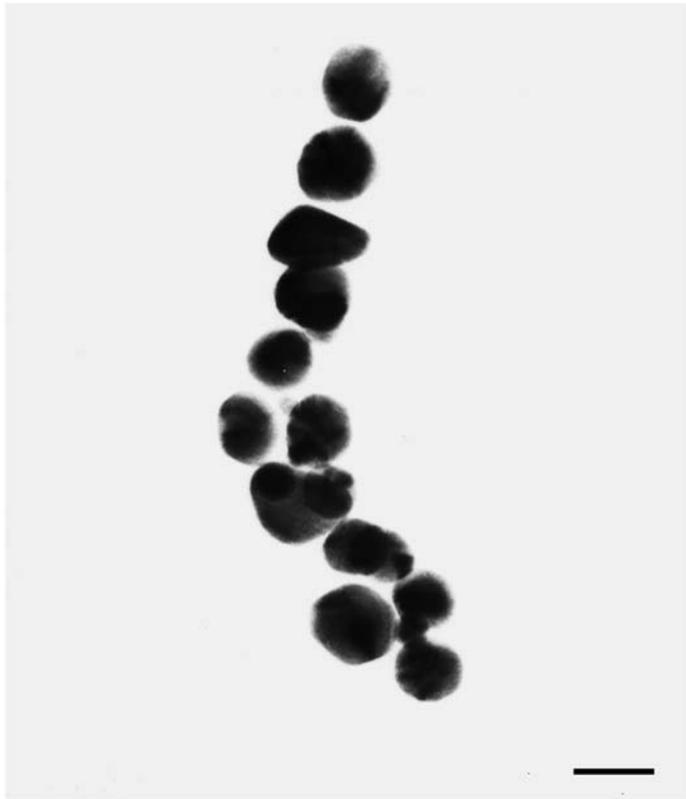

Figure_3

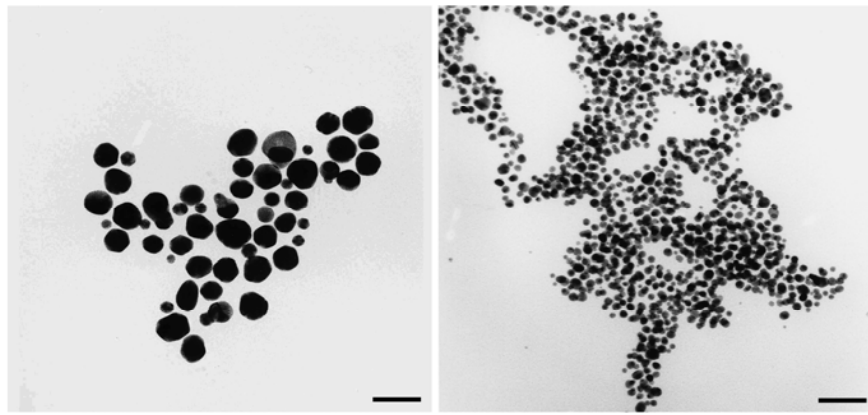

Figure_4

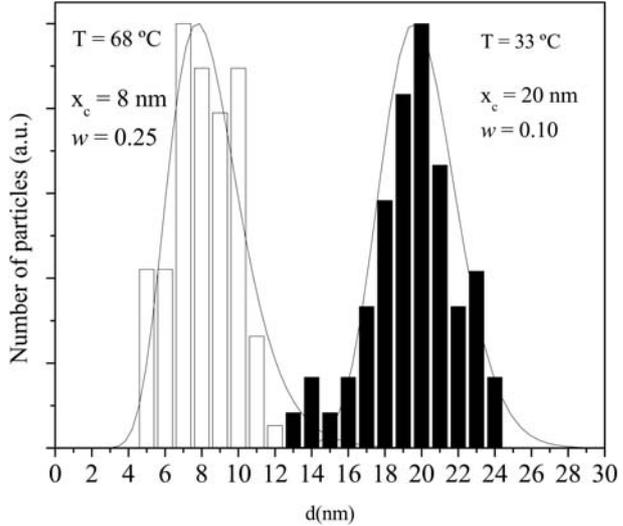

Figure_5

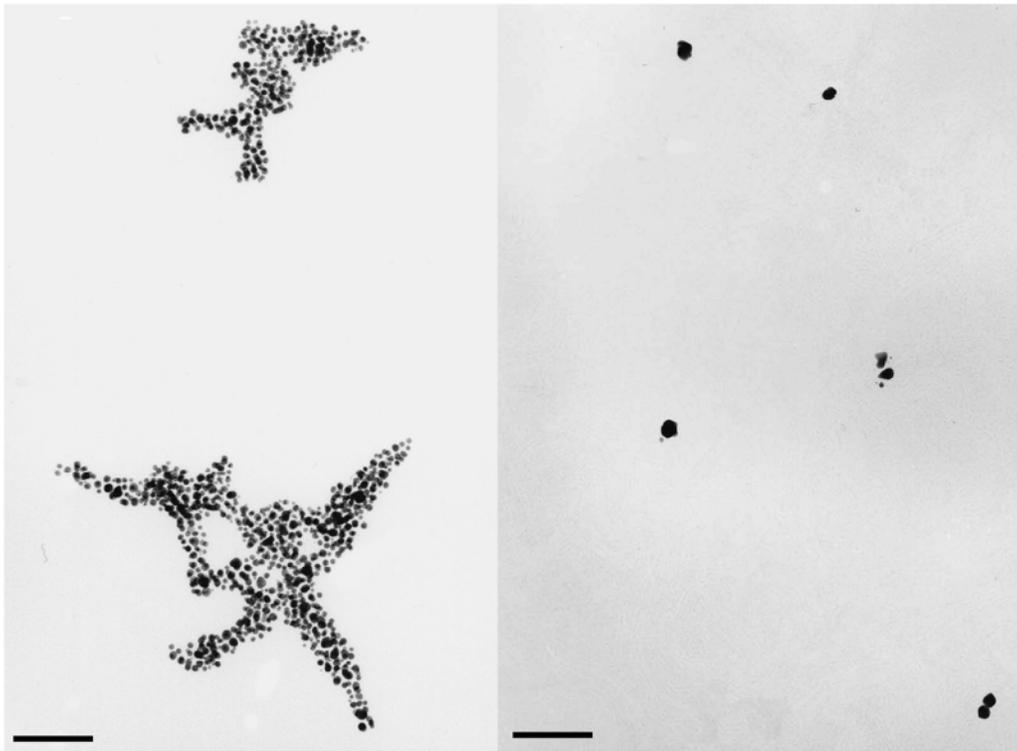

Figure_6

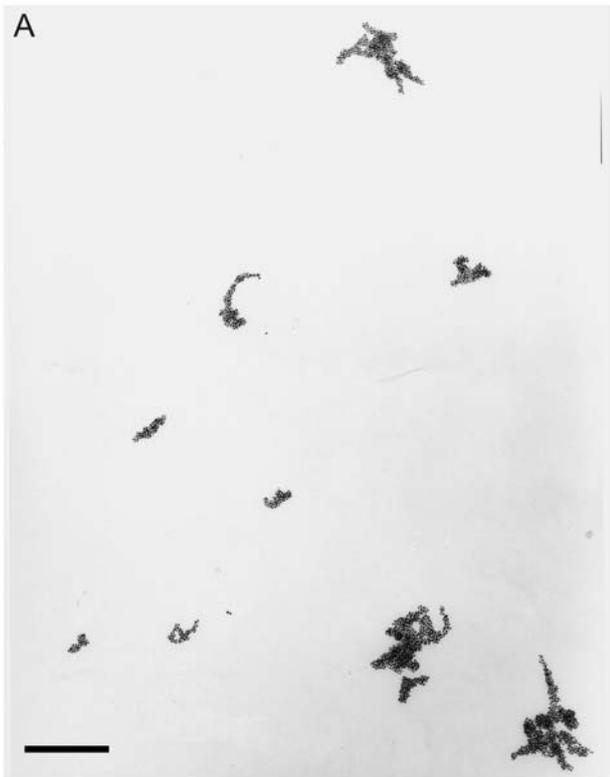 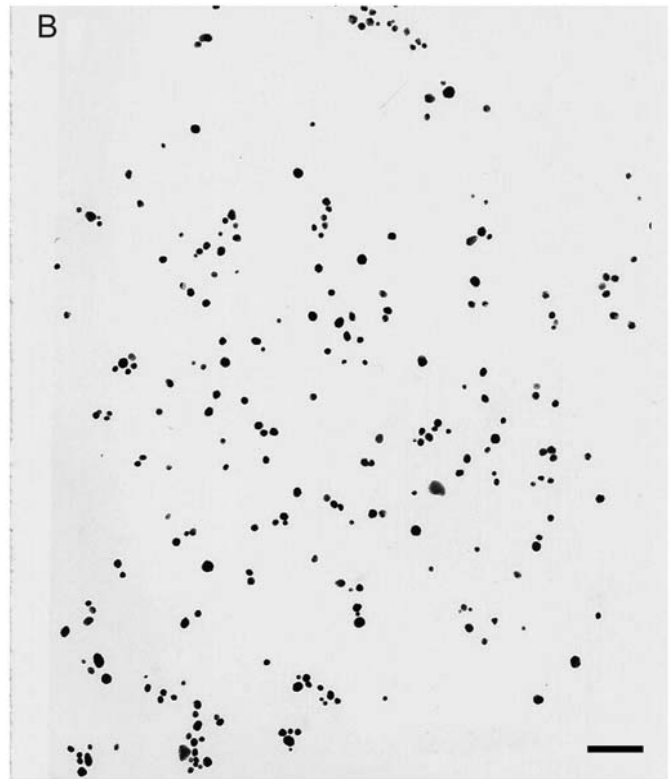

Figure_7

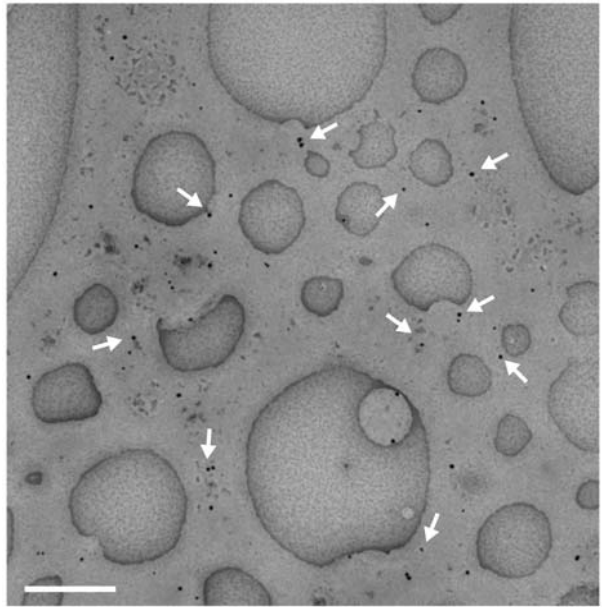

Figure_8

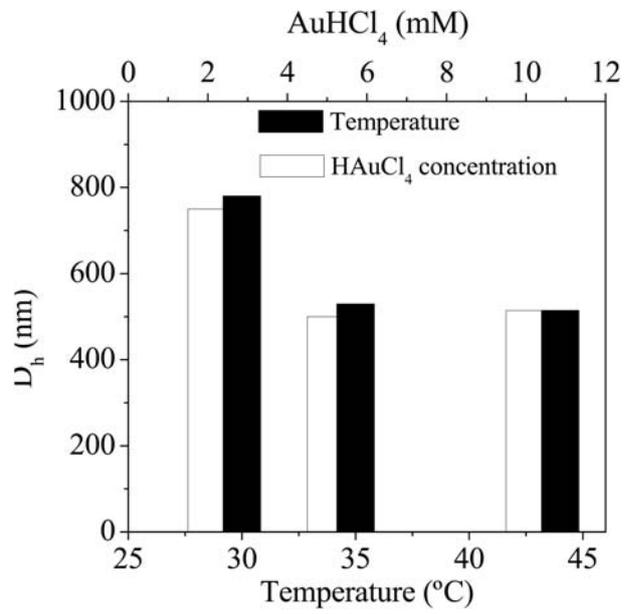

Figure_9

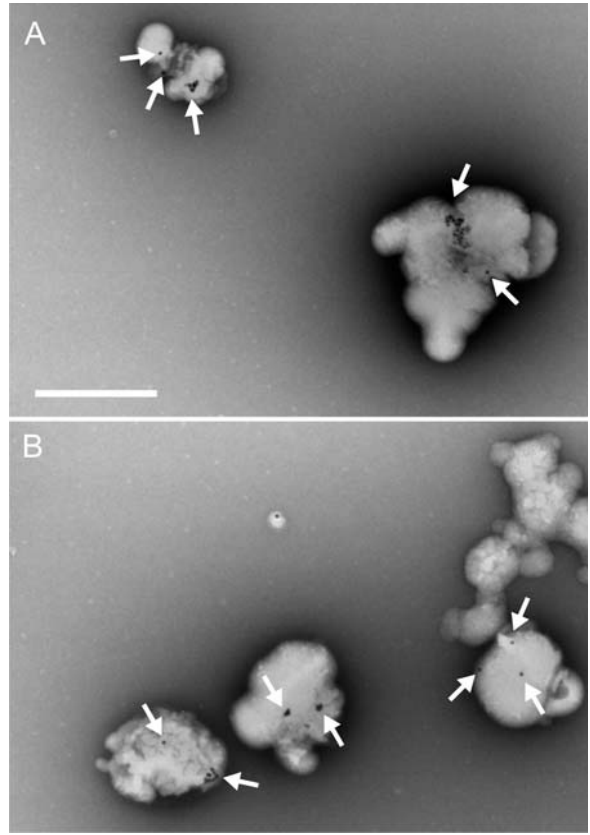

Figure_10

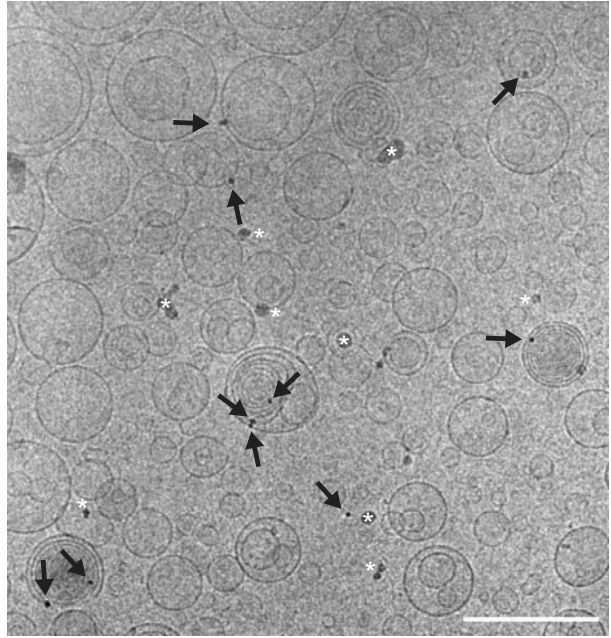

Figure_11

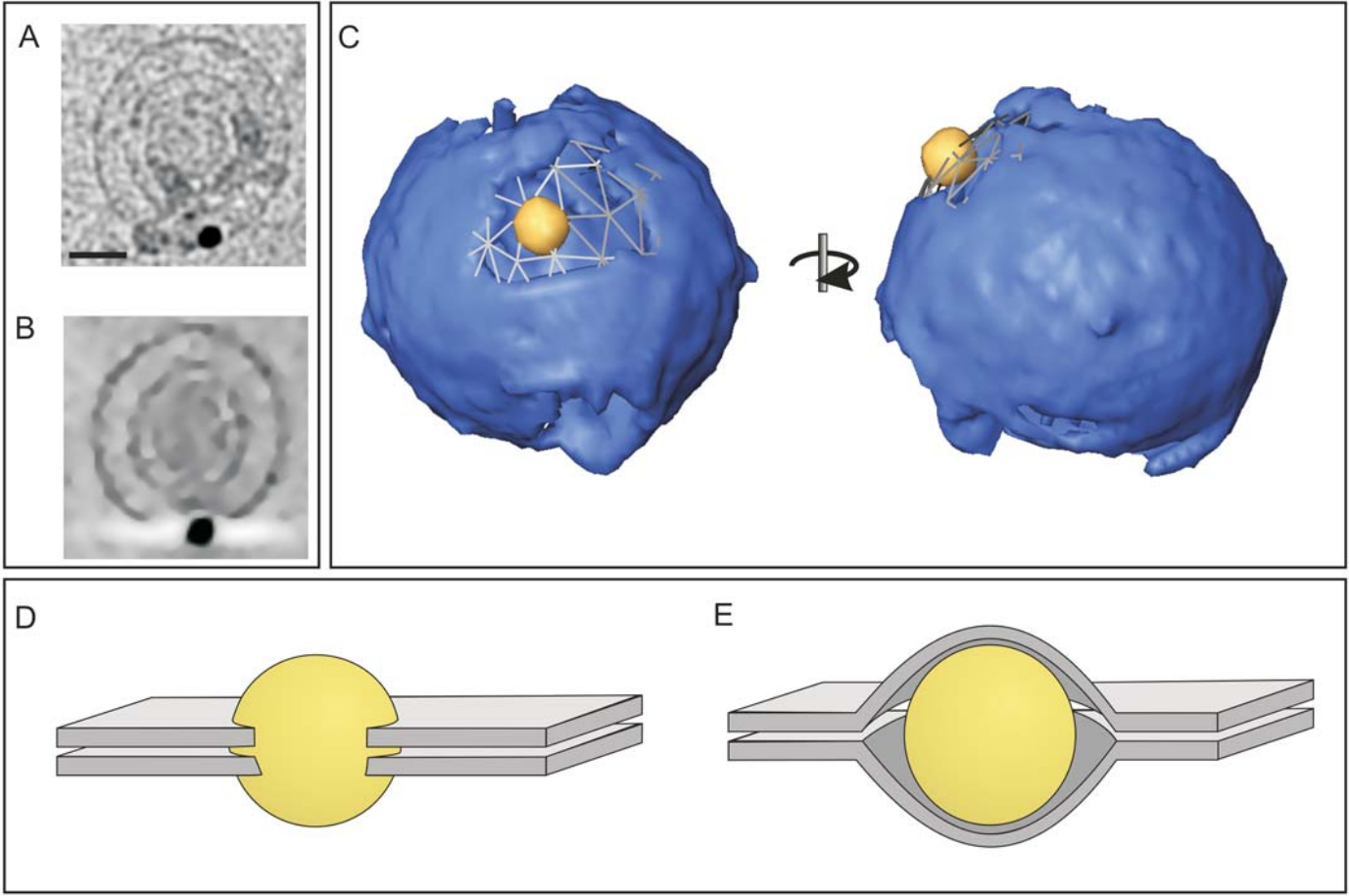

# SI_Figure_1

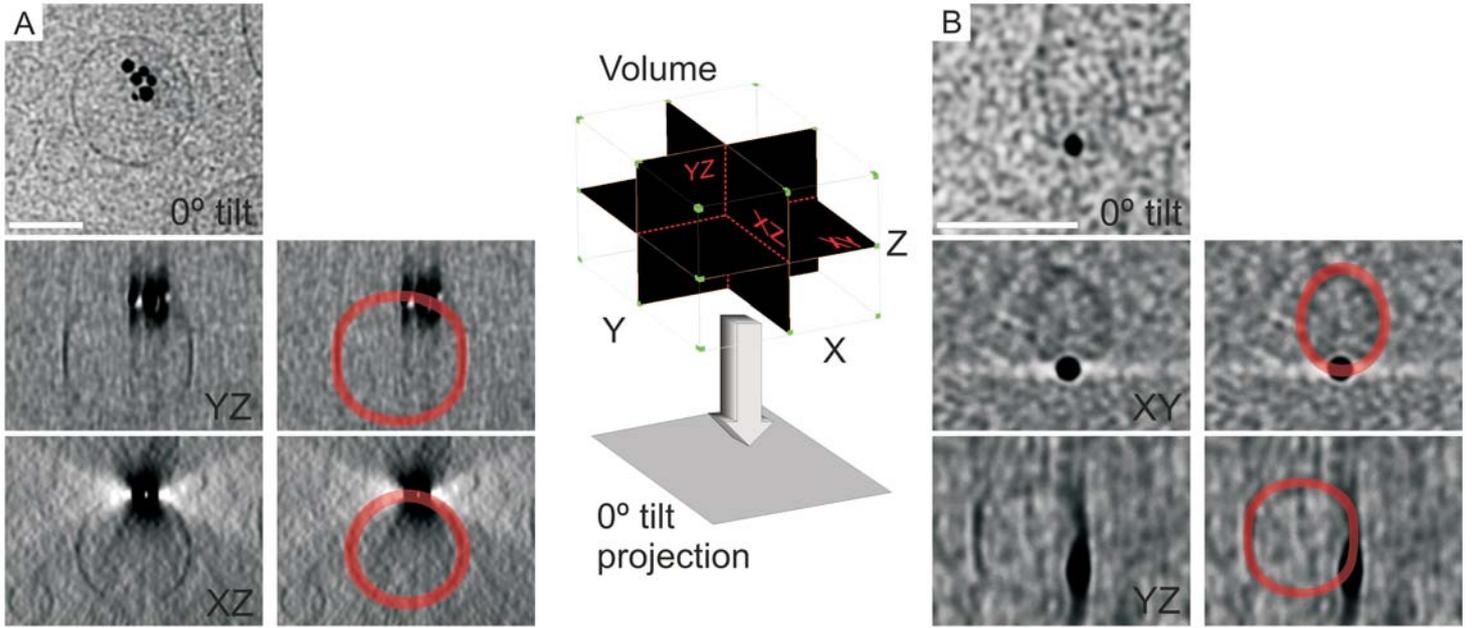